\documentclass[manuscript]{aastex631}
\submitjournal{{\it Astrophys. J. Lett.}}

\shorttitle{SEP Track Production Rates}
\shortauthors{Poppe et al.}
\graphicspath{{./}{figures/}}
\begin{document}

\title{Solar-Energetic-Particle Track-Production Rates at 1 au: \\ Comparing In-Situ Particle Fluxes with Lunar Sample-Derived Track Densities}

\correspondingauthor{A. R. Poppe}
\email{poppe@berkeley.edu}

\author[0000-0001-8137-8176]{A. R. Poppe}
\affiliation{Space Sciences Laboratory, University of California at Berkeley, Berkeley, CA, 94720}

\author[0000-0002-7478-7999]{P. S. Szabo}
\affiliation{Space Sciences Laboratory, University of California at Berkeley, Berkeley, CA, 94720}

\author{E. R. Imata}
\affiliation{Dept. of Astronomy, University of California at Berkeley, Berkeley, CA, 94720}

\author[0000-0003-1560-2939]{L. P. Keller}
\affiliation{NASA Johnson Space Center, Mail Code XI3, Houston, Texas 77058, USA}

\author{R. Christoffersen}
\affiliation{Jacobs, NASA Johnson Space Center, Mail Code X13, Houston, Texas 77058, USA}

\begin{abstract}

Heavy ($Z>26$) solar energetic particles (SEPs) with energies $\sim$1 MeV/nucleon are known to leave visible damage tracks in meteoritic materials.
The density of such `solar flare tracks' in lunar and asteroidal samples has been used as a measure of a sample's exposure time to space, yielding critical information on planetary space weathering rates, the dynamics and lifetimes of interplanetary dust grains, and the long-term history of solar particle fluxes.
Knowledge of the SEP track accumulation rate in planetary materials at 1 au is critical for properly interpreting observed track densities.
Here, we use in-situ particle observations of the 0.50$-$3.0 MeV/nuc Fe-group SEP flux taken by NASA's Advanced Composition Explorer (ACE) to calculate a flux of track-inducing particles at 1 au of $6.0\times10^5$ cm$^{-2}$ yr$^{-1}$ str$^{-1}$.
Using the observed energy spectrum of Fe-group SEPs, we find that the depth distribution of SEP-induced damage tracks inferred from ACE measurements matches closely to that recently measured in lunar sample 64455; however, the magnitude of the ACE-inferred rate is approximately 25$\times$ higher than that observed in the lunar sample.
We discuss several hypotheses for the nature of this discrepancy, including inefficiencies in track formation, thermal annealing of lunar samples, erosion via space weathering processing, and variations in the SEP flux at the Moon, yet find no satisfactory explanation.
We encourage further research on both the nature of SEP track formation in meteoritic materials and the flux of Fe-group SEPs at the lunar surface in recent and geologic times to resolve this discrepancy.

\end{abstract}



\section{Introduction}
\label{intro}

Objects exposed to the harshness of space are subjected to a wide range of charged-particle irradiation that can physically and chemically alter their nature.
In particular, fluxes of $\sim$1 MeV/nuc, high-Z ($Z>26$, typically Fe and heavier) solar energetic particles (SEPs) have been shown to leave observable damage tracks in meteoritic minerals, including interplanetary dust grains \citep[e.g.,][]{Bradley_1984, Thiel_1991}, meteorites \citep[e.g.,][]{Goswami_1981}, and returned lunar and asteroidal samples \citep[e.g.,][]{Crozaz_1972, Blanford_1974, Keller_2014}.
The characterization of these tracks, including their overall density as well as their depth profiles, informs us about both the exposure age of planetary materials to space \citep[e.g.,][]{Bradley_1984, Sandford_1986, Keller_2021, Keller_2022} and the solar energetic particle flux over solar system timescales \citep[e.g.,][]{Price_1970, Zinner_1980}.

A key question in such studies is the rate at which typical meteoritic minerals accumulate SEP tracks at 1 au.
% - lab measurements of 64455 and applications to EKB dust at 1 au
\citet{Blanford_1974} used acid-etching techniques on Apollo 16 sample 64455 to determine an SEP track accumulation rate of $\sim$$6\times10^5$ tracks cm$^{-2}$ yr$^{-1}$ for an assumed 2$\pi$ exposure; however, this analysis required a series of renormalizations and extrapolations, which leaves uncertainty as to the robustness of the final results.
Recently, laboratory measurements of SEP-induced tracks within lunar sample 64455 using more advanced techniques have yielded a re-calibration of the rate of SEP track formation in minerals at 1 au of $4.4\pm0.4\times10^4$ tracks cm$^{-2}$ yr$^{-1}$, again assuming a 2$\pi$ exposure \citep{Keller_2021}.
In turn, the SEP track-formation rate determined in \citet{Keller_2021} has led to the conclusion that interplanetary dust grains collected from the terrestrial stratosphere with unusually high track densities ($\gtrsim$$10^{11}$ tracks cm$^{-2}$) may originate from the Edgeworth-Kuiper Belt beyond Neptune \citep{Keller_2022}.
Such a conclusion has significant implications for the distribution and dynamics of interplanetary dust grains throughout the solar system \citep[e.g.,][]{Liou_1999, Kuchner_2010, Poppe_2019b}, yet such conclusions rely critically on knowledge of the SEP track accumulation rate.

Here, we use a complementary approach to calculating the track-inducing flux of SEPs at 1 au via in-situ observations from NASA's Advanced Composition Explorer (ACE) \citep{Stone_1998}, which has been in a heliocentric orbit at the solar-terrestrial Lagrange-1 point since 1998. 
We compare this in-situ derived rate to the sample-derived track-formation rate of \citet{Keller_2021} and find that while the shape of the track density versus depth profile matches the sample data well, the overall magnitude of the in-situ derived rate is approximately 25$\times$ higher than the lunar sample-derived rate.
We assess several possibilities for the discrepancy between these two measurement approaches, yet find no obvious explanation and therefore urge additional laboratory and in-situ experiments on the nature of SEP track accumulation in meteoritic and lunar minerals.

~

%--------------------------------------------------------
\section{Fe-Group Solar Energetic Particle Flux at 1 au}
\label{sec2}

To calculate the flux of SEP track-producing particles at 1 au, we use observations taken by the Ultra-Low-Energy Isotope Spectrometer (ULEIS) instrument onboard NASA's Advanced Composition Explorer. 
Launched in 1997, the ACE mission was designed to measure the elemental and isotopic composition of space-based particles over a wide range of energies ($\sim$keV/nuc to $\sim$GeV/nuc) and masses (atomic numbers, 1 $\le Z \le$ 28) \citep{Stone_1998}.
Amongst a broader payload, the ULEIS instrument measures the compositionally resolved energy spectra of elements between He (Z=2) and Ni (Z=28) in the energy range, $\sim$45 keV/nuc $< E <$ $\sim$few MeV/nuc \citep{Mason_1998}.
Solar-flare track production within meteoritic materials only occurs for very heavy nuclei with Z$\ge \sim 26$ \citep[e.g., Ch. 1,][and refs. therein]{Fleischer_1975}; thus, we focus our analysis on the Fe-group (Z$\ge$26) ions measured by ULEIS.
We acquired the full dataset of Fe-group flux measured by ULEIS between 1998 and mid-2023 in the energy range, 0.035 $< E <$ 3.07 MeV/nuc, via NASA's Coordinated Data Analysis Website (CDAWeb).
Note that while the Fe-group flux reported by ULEIS technically includes all species with $Z > 26$, the elemental abundance of minor species in the solar wind in this range is dominated by Fe (Z=26) \citep[e.g.,][]{Meyer_1985, Bochsler_1987}.
We also note that prior to mid-2001, the ULEIS data occasionally suffered from saturated count rates for the largest SEP events [{\it G. Mason, priv. comm.}, 2023]; thus, we restrict our analysis to the $\sim$21-year time period 2002$-$2023.

\begin{figure}[htb]
	\includegraphics*[width=6in]{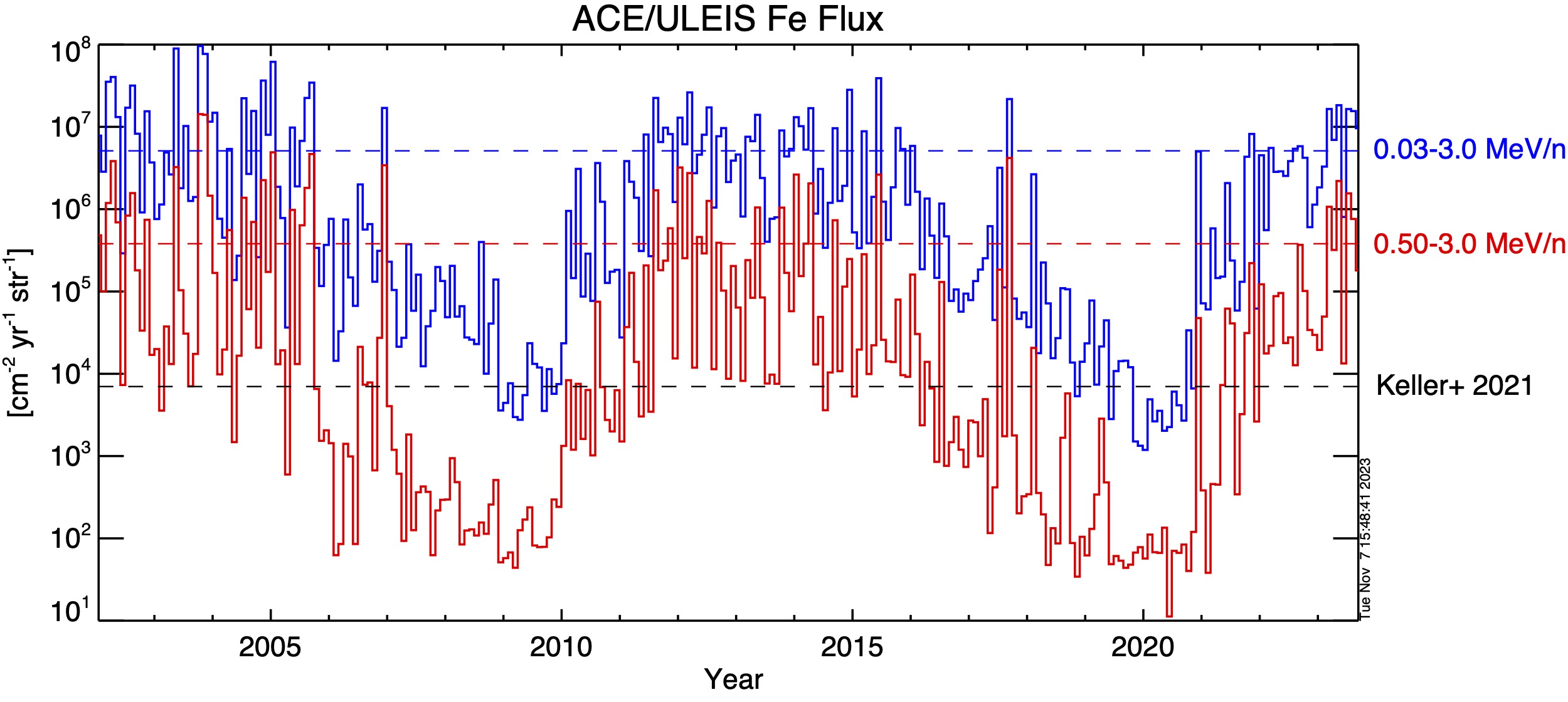}
	\caption{The monthly averaged Fe SEP flux measured by ACE/ULEIS for two energy ranges: (blue) 0.03 $-$ 3.0 MeV/nuc and (red) 0.50 $-$ 3.0 MeV/nuc. Average values for each separate energy range are shown as dashed lines. The SEP track formation flux at 1 au inferred from \citet{Keller_2021} is shown as the black dashed line.}
	\label{uleis_timeseries}
\end{figure}

Figure \ref{uleis_timeseries} shows the monthly averaged flux of Fe-group SEPs from 2002 to 2023 over two different energy ranges: (i) the full energy range measured by ULEIS, 0.035 $< E <$ 3.07 MeV/nuc, and (ii) the approximate energy range within which Fe-group SEPs are expected to generate observable tracks, 0.50 $< E <$ 3.07 MeV/nuc \citep[discussed below; see also][]{Szenes_2010}.
Note that Fe-group SEPs with energies greater than this range will produce tracks deeper within a material once they have shed sufficient excess energy and thus, could also contribute to track densities; however, the steep slope of the energy distribution (discussed below) implies that the exclusion of such higher-energy particles does not overly affect our results.
Both curves are similar in shape, displaying both short-term variation due to individual impulsive CMEs and/or solar flares and long-term variation corresponding to the 11-year solar cycle for solar cycles 23, 24, and the beginning of solar cycle 25.
For both curves, the respective horizontal dotted lines denote the mean flux over this time range, specifically $3.2\times10^6$ cm$^{-2}$ yr$^{-1}$ str$^{-1}$ for the full energy range and $3.8\times10^5$ cm$^{-2}$ yr$^{-1}$ str$^{-1}$ for the $E>0.50$ MeV/nuc range.
Figure \ref{uleis_espec} shows the differential flux as a function of energy-per-nucleon for Fe-group SEPs observed by ULEIS averaged over the full time period presented in Figure \ref{uleis_timeseries}.
As shown by the fitted curve, the differential spectrum is well described by a power law, $J_{Fe}(E) = 2.3\times10^5 \cdot E^{-1.70}$ cm$^{-2}$ yr$^{-1}$ str$^{-1}$ (MeV/nuc)$^{-1}$.
Based on an analysis of lunar sample 64455, \citet{Blanford_1974} found that a long-term-averaged SEP spectral slope of $\gamma$ = -1.9 was consistent with the observed solar flare track density distribution versus depth.
This spectral slope is slightly steeper than that measured by ACE ($\gamma = -1.70$), but within reason given the different observational approaches.

We also verified the differential Fe-group flux measured by ACE by comparison to concurrent Fe-group measurements in a slightly lower energy range of 0.03$-$0.5 MeV/nuc by the Supra-Thermal Energetic Particle (STEP) subsystem on the Energetic Particle: Acceleration, Composition, and Transport (EPACT) investigation on the Wind spacecraft \citep{vonRosenvinge_1995}.
Within quoted energy resolution and error bars, the differential Fe-group flux measured by Wind/STEP matches that reported by ACE.

% ----------------------------------------------------------
\section{Inferring Track Production Rates at 1 au}
\label{sec3}

\begin{figure}[htb]
	\includegraphics*[width=6in]{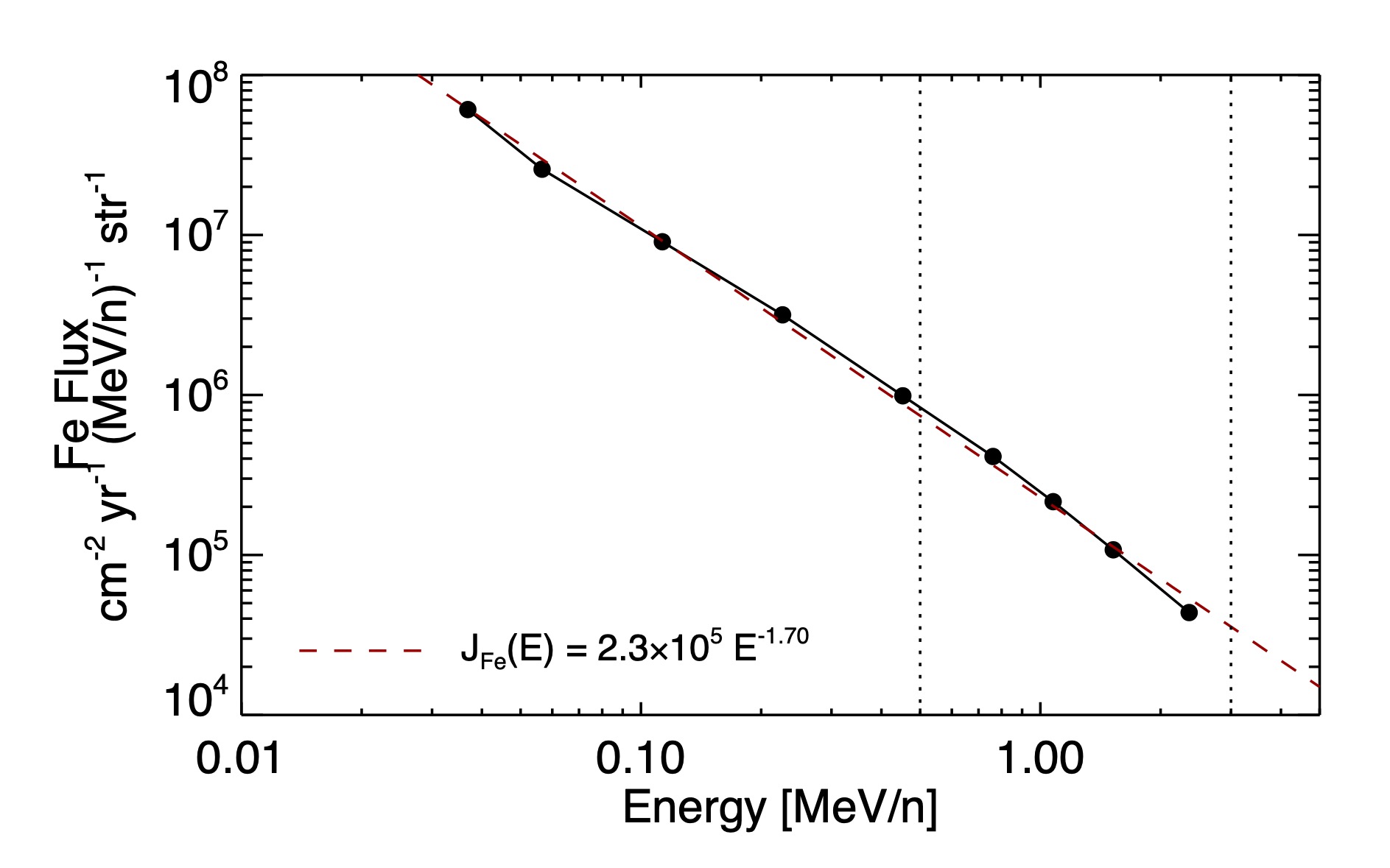}
	\caption{The differential energy spectrum of Fe SEPs measured by ACE/ULEIS between 0.03 MeV/nuc and 3.0 MeV/nuc. The best-fit power law spectrum is denoted by the dashed red line. The approximate energy range in which Fe-group SEPs leave damage tracks in meteoritic materials is denoted by the vertical dotted lines.}
	\label{uleis_espec}
\end{figure}

\begin{figure}[htb]
	\includegraphics*[width=6in]{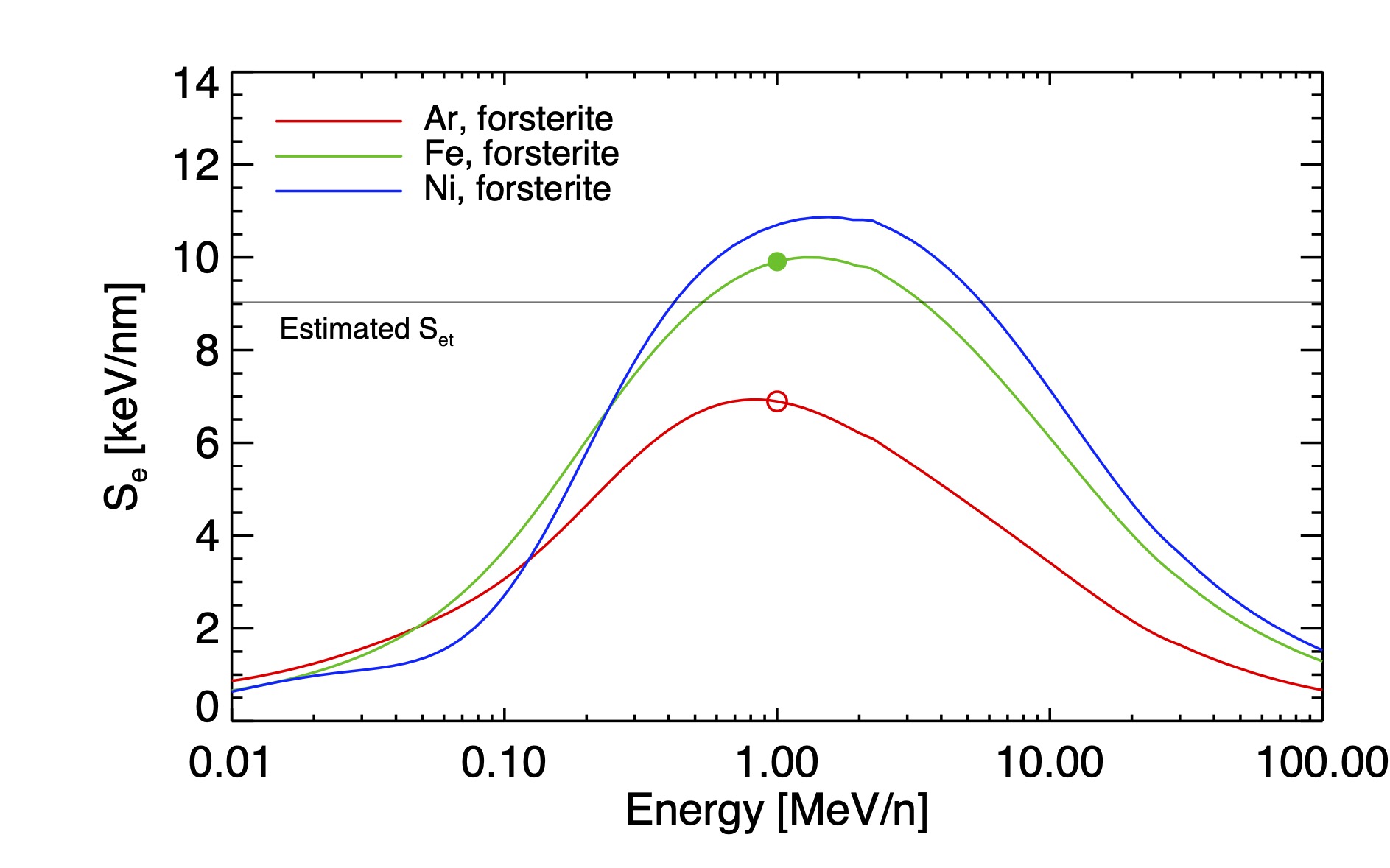}
	\caption{The electronic stopping power, $S_e$, for three incident ion species (Ar, red; Fe, green; Ni, blue) in a forsterite mineral. The green closed circle and red open circle represent experimental measurements by \citet{Szenes_2010} that did and did not register tracks, respectively. Correspondingly, the minimum required $S_e$ for track formation estimated by \citet{Szenes_2010} is shown as the horizontal line.}
	\label{Se_enstatite}
\end{figure}

Using the time-averaged Fe-group SEP flux measured by ACE, we employ a simple analytical model to calculate the SEP-induced track density as a function of depth in lunar and/or meteoritic materials at 1 au.
We obtained the electronic stopping power as a function of energy for Fe incident on an forsterite grain (Mg:Si:Fe:O = 27:12:4:56; matching that of \citet{Szenes_2010}) from the TRIM.SP code \citep{Ziegler_2010}, shown in Figure \ref{Se_enstatite}.
In this energy range (E $>$ 0.01 MeV/nuc), the electronic stopping power dominates over the nuclear stopping power and peaks near 1.5 MeV/nuc.
Previous laboratory work has shown that track formation in insulators occurs only when incident particles deposit energy above a given linear energy density threshold.
Using a forsterite sample, \citet{Szenes_2010} have shown that 56 MeV Fe (1.0 MeV/nuc) ions leave tracks with nearly unit efficiency, while 48 MeV Ar (1.2 MeV/nuc) ions do not register any tracks.
The 1.0 MeV/nuc Fe ions have a peak electronic stopping power of 9.9 keV/nm (green line, Figure \ref{Se_enstatite}) while the 1.2 MeV/nuc Ar ions have an electronic stopping power of 6.9 keV/nm (red line, Figure \ref{Se_enstatite}).
\citet{Szenes_2010} further present an analytical formula for the threshold electronic stopping power, $S_{et}$, above which particles will induce track formation and below which, they will not.
From their experiments, \citet{Szenes_2010} derive a threshold value, $S_{et} = 9.04$ keV/nm (horizontal line, Figure \ref{Se_enstatite}), consistent with the registration of tracks from 1.0 MeV/nuc Fe but not 1.2 MeV/nuc Ar.
Adopting this threshold, we estimate that Fe SEPs must fall within an energy range, $0.50 < E < 3.2$ MeV/nuc, in order to register track formation within forsterite minerals.
Note that other minerals will have slightly different electronic stopping powers and thus, slightly different energy ranges to which they are susceptible to SEP track formation.
We also note that experimental and computational studies have shown that ions with energies on opposite sides of the Bragg peak have different electronic stopping power thresholds for track formation \citep[the so-called `velocity effect'; e.g.,][]{Constantini_1992, Szenes_2010, Rymzhanov_2019} which could affect the overall energy range for track formation. 
These experiments have also shown that the effect is primarily manifested as higher electronic stopping power thresholds (i.e., reduced track formation rates) at energies {\it above} the Bragg peak. However, considering the steep slope of the differential flux shown in Figure \ref{uleis_espec}, use of a constant $S_{et}$ as opposed to a non-linear threshold that takes into account the velocity effect is likely to have only a minor effect on the overall track production rate calculated here.

In the analytic model, we calculate the track production rate, $d\rho/dt$, as a function of depth, $z$, by integrating the incident Fe-group SEP flux via, 
\begin{equation}
\frac{d\rho(z)}{dt} = \pi \int_{E_{min}(z)}^{E_{max}(z)} J_{Fe}(E,z=0)~dE,
\end{equation}
where $J_{Fe}(E,z=0)$ is the differential Fe SEP flux at the surface of the grain as derived above and shown in Figure \ref{uleis_espec}, [$E_{min}(z)$, $E_{max}(z)$] are the minimum and maximum energies of the upstream distribution that are capable of registering tracks at depth $z$, and the factor of $\pi$ accounts for the exposed solid angle of a point on the lunar surface \citep[see also][]{Fraundorf_1980}.
To determine [$E_{min}(z)$, $E_{max}(z)$], we numerically integrated the penetration of Fe SEPs into the mineral surface using the electronic stopping power shown in Figure \ref{Se_enstatite}.
This step allows the model to correctly account for SEPs that are initially above the 3.2 MeV/nuc threshold, yet begin to produce tracks at greater depths once they have shed sufficient energy to fall within the $0.50 < E < 3.2$ MeV/nuc range.
For simplicity, we assume all SEPs to be normally incident to the surface.
Finally, to compare with the results of \citet{Keller_2021}, who measured the track density as a function of depth for the 2 Myr-exposed lunar rock 64455, we multiplied $d\rho(z)/dt$ by $2\times10^6$ yr to obtain the track density versus depth, $\rho(z)$.

\begin{figure}[htb]
	\includegraphics*[width=6in]{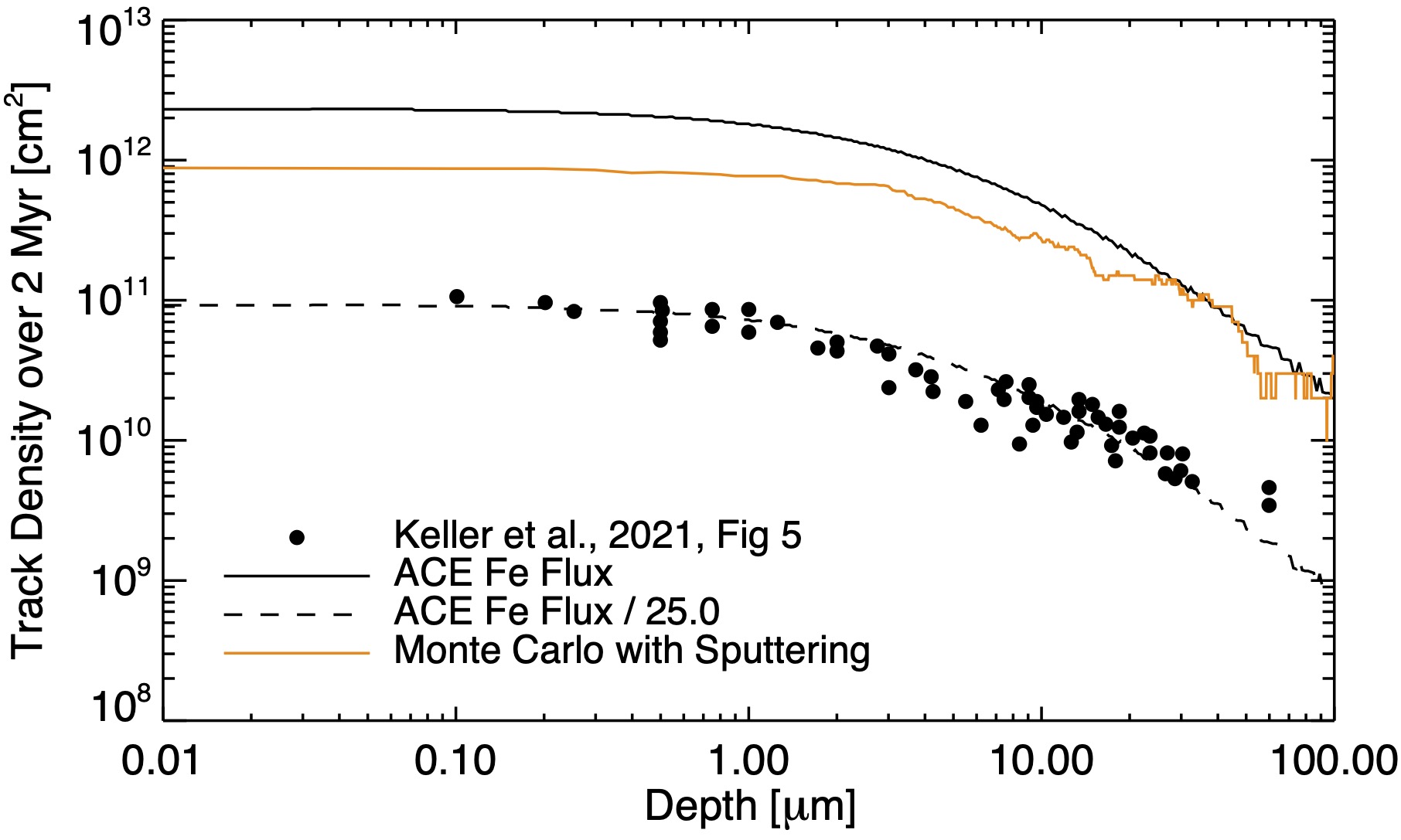}
	\caption{The predicted track density as a function of depth for an objects exposed for 2 Myr at 1 au as determined by the ACE in-situ measurements (solid line). Solid dots reproduce the measurements of \citet{Keller_2021} along with the ACE-predicted flux lowered by a factor of 25 (dashed line). The orange line is the result of a Monte Carlo model for track formation taking into account a charged-particle erosion rate of 7 $\mu$m/Myr.}
	\label{compare}
\end{figure}

Figure \ref{compare} compares the analytic derivation for $\rho(z)$ described above and the data reported from \citet{Keller_2021}.
The analytical track density calculation based on the ACE-measured Fe SEP flux yields a maximum track density at the surface ($z = 0.01$ $\mu$m) of $2.8\times10^{12}$ cm$^{-2}$ with a gradual decrease as a function of depth.
At 100 $\mu$m depth, the track density has fallen to approximately $3\times10^{11}$ cm$^{-2}$.
Comparing to the \citet{Keller_2021} results, the ACE-calculated track density has a nearly identical shape with respect to depth, but is $\sim25\times$ higher; the dashed curve denotes the ACE-calculated flux divided by 25 to illustrate this comparison.
To first order then, the track production rates derived from in-situ Fe-group SEP measurements are in conflict with sample-derived track production rates reported in \citet{Keller_2021}. 
Below, we discuss possible reasons for this discrepancy.

% -----------------------------------------
\section{Discussion}
\label{sec4}

The overestimation of the in-situ particle flux-derived track density derived from ACE relative to the lunar sample-derived track density suggests that some process is acting to either suppress track formation (relative to our current understanding of track formation) or erase tracks at some rate after they have formed.
Here, we discuss several possible hypotheses that could account for such an effect, including (i) variations in the efficiency of SEP-induced track registration within meteoritic minerals, (ii) thermal annealing of tracks, (iii) grain and track erosion processes, (iv) shielding of SEP fluxes locally at the Moon compared to L1, (v) long-term variations in the SEP flux at 1 au, and (vi) uncertainties in track-density measurement techniques; however, we note that each of these hypotheses suffers in some critical way and a clear resolution is not yet in hand.

\subsection{Track Registration Efficiency}

Our calculations of track production rates based on in-situ observed particle fluxes require knowledge of the threshold electronic stopping power required for track registration \citep[e.g.,][]{Szenes_2010}, which is likely to vary across different minerals. 
Thus, changes in the assumed threshold could impact the total track production rate. 
To explore this, we repeated our calculations in Equation 1 using the same input Fe SEP flux but with progressively higher electronic stopping power thresholds (i.e., implying a less sensitive mineral for track formation).
We found that the 25$\times$ lower track production rate could only be achieved if the electronic stopping power threshold was increased to nearly the maximum observed (i.e., 99.95\% of the maximum), such that Fe SEPs only induced track formation over an incredibly narrow range of energies ($\approx$1.34$-$1.53 MeV/nuc).
We consider such ``fine-tuning" of the electronic stopping power threshold to be unrealistic, in particular in the face of significant experimental evidence that SEP Fe ions over a broader range of energies can induce track formation with unit efficiency \citep[e.g.,][]{Fleischer_1965, Seitz_1970, Price_1973, Szenes_2010}.
Additionally, such a narrow energy range for track formation would lead to the formation of exceedingly short tracks ($\sim$20 nm); however, track lengths many tens of microns are routinely observed in space-exposed minerals \citep[e.g.,][]{Blanford_1974, Bull_1975, Keller_2021}.
Nevertheless, additional laboratory measurements that methodically characterize the track registration efficiency in a variety of minerals over a broad range of incident energies could help to better elucidate the exact energy range within which track formation occurs. 

\subsection{Track Annealing}

SEP tracks within materials can be annealed via exposure to high temperatures, which promotes atomic mobility within the crystal lattice.
Early work by \citet{Price_1973} suggested that at maximum lunar surface temperatures ($\sim$130 $^{\circ}$C), thermal annealing of SEP-induced damage tracks could be effective on timescales of $\sim$10$^5$$-$10$^6$ years (see their Figure 9), which could plausibly affect the comparison between ACE-measured and lunar-derived SEP track densities.
However, the suggestion by \citet{Price_1973} relied on extrapolation of annealing at much higher temperatures and shorter timescales and other experiments have not supported this.
Tracks in most minerals do not show appreciable annealing for temperatures below $\sim$400 $^{\circ}$C \citep[e.g.,][]{Bull_1975, Afra_2014}, which is far above temperatures encountered on the lunar surface.
Furthermore, as discussed in e.g., \citet{Paul_1992}, tracks undergoing annealing typically display a characteristic behavior in which a single continuous track develops gaps along its axis as individual portions of the track anneal (see their Figure 6).
However, no such `gapped' tracks indicative of thermal annealing have been reported in lunar sample 64455 \citep{Keller_2021}, suggesting that annealing of lunar samples$-$even on geologic timescales$-$is not occurring.

\subsection{Grain Erosion Mechanisms}

Regolith grains exposed to space are subject to erosive processes, chief among which is sputtering of individual atoms via incident charged particles \citep[e.g.,][]{Biersack_1984, Szabo_2018}.
Decades of laboratory measurements have quantified the sputtering yield of silicate surfaces subject to ion bombardment in the keV energy range \citep[e.g.,][]{Biersack_1984}.
Using typical values for the solar wind flux at 1 au and the combined proton and alpha sputtering yield, grains at 1 au are eroded via charged-particle sputtering at a rate of $\sim$7 $\mu$m/Myr.
Thus, over the 2 Myr exposure of lunar sample 64455, we would expect $\sim$14 $\mu$m of erosion.
To account for this erosion rate in the accumulation of tracks, we developed a simple Monte Carlo model whereby tracks are numerically created within a model grain with a depth profile determined from the ACE measurements as shown in  Figure \ref{uleis_espec} and simultaneously eroded from the top down (i.e., from the exposed grain surface) at the 7 $\mu$m/Myr sputtering rate. 
After $\approx$1.3 Myr, the track density versus depth profile reached an equilibrium, shown in Figure \ref{compare} as the orange curve.
Even when accounting for charged-particle sputtering, the track density at the grain surface is $\sim$1.3$\times10^{12}$ cm$^{-2}$, lower than the value without sputtering, 2.8$\times10^{12}$ cm$^{-2}$, yet still a factor of $\sim$14 higher than that measured by \citet{Keller_2021}.
Thus, charged-particle sputtering, while likely reducing the track density somewhat, is insufficiently intense to explain the observed discrepancy between ACE and lunar sample 64455.

\subsection{SEP Shielding at the Moon}

Discrepancies between the SEP flux measured by ACE at the Earth-Sun Lagrange 1 point and the Moon could in theory arise due to local shielding of the lunar surface from SEPs.
Remanent crustal magnetic fields are widespread across the lunar surface, with magnitudes up to at least hundreds of nanotesla \citep[e.g.,][]{Mitchell_2008}.
In-situ particle measurements have shown that some crustal fields are of sufficient strength and coherency to reflect keV-energy solar wind protons \citep[e.g.,][]{Lue_2011, Saito_2012, Poppe_2017} likely due to the formation of quasi-static electric fields within the anomaly interaction regions \citep[e.g.,][]{Fatemi_2015, Deca_2015}.
At MeV energies, however, neither magnetic fields nor quasi-static electric fields are thought capable of reflecting particles.
MeV-scale electric fields are exceedingly unlikely to exist within such anomalies and a 1 MeV/nuc $^{56}$Fe$^{20+}$ SEP in the presence of a 1000 nT field has a gyroradius of $\sim$400 km, which is much larger than the coherency scale of most magnetic anomalies.
Thus, the presence of lunar crustal magnetic fields are unlikely to provide any shielding to lunar soil from 1 MeV/nuc Fe-group SEPs.

An additional source of discrepancy between L1-measured SEP fluxes and those at the lunar surface could come from the Moon's transit through the terrestrial magnetotail each lunation; however, this is also unlikely for two reasons.
First, the Moon only spends approximately one quarter of its orbit in the magnetotail which plainly cannot account for the factor of 25 difference discussed above.
Furthermore, recent analysis of in-situ particle measurements at the Moon have shown that SEPs likely have broad access to the lunar environment even within the terrestrial magnetotail due to the `open' nature of magnetotail lobe field lines to the solar wind \citep{Liuzzo_2023}.
Finally, shielding of specific locations on the lunar surface by the solid obstacle of the Moon itself, while highly effective at keV energies \citep[e.g.,][]{Fatemi_2012}, appears to yield only small or even negligible results at MeV energies \citep[e.g.,][]{XuX_2017}.
Nevertheless, in-situ SEP measurements placed at the lunar surface, both on the nearside and farside for comparison, could help to better constrain any local or regional shielding effects.

% ----------------------------------
\subsection{Long-term SEP Variability}

We also consider the possibility that the in-situ measurements from the ACE spacecraft during the modern space age are not representative of the 2 Myr-averaged SEP flux presumably recorded by sample 64455.
In other words, was lunar sample 64455 exposed on the lunar surface during an extended solar minimum after which the modern age is a kind of grand maximum in solar activity?
While variations over multiple time scales in solar and stellar behavior are a well-documented phenomenon \citep[e.g.,][]{Usoskin_2023} and solar cycles 17$-$23 ($\sim$1940-2000) are considered a `Modern Maximum', both sunspot measurements over the past $\sim$300 years \citep[e.g.,][]{Usoskin_2016b, Muscheler_2016, Carrasco_2016} and cosmogenic radionuclide data over the past several millenia \citep[e.g.,][]{McCracken_2013, Usoskin_2016} do not suggest that the current space-age measurements are exceedingly atypical.
That acknowledged, the current available history of solar activity ($\sim$10$^4$ yrs) falls well short of characterizing the $2\times10^6$-year exposure age of lunar sample 64455 and thus, does not entirely rebut the question.
Nevertheless, the idea that the recent 10,000 years are representative of an extreme maximum nearly 25$\times$ higher than the million-year average is not particularly tenable and thus, we adopt the position that$-$at least to first order$-$the modern space-age measurements taken by ACE are reasonably representative of the past two million years.

\subsection{Uncertainties in Track-Density Measurement Techniques}

As noted in the Introduction, recent TEM measurements of SEP-induced tracks in meteoritic materials by \citet{Keller_2021} have revised the sample-based track accumulation rate at 1 au downwards by a factor of $\sim$20 relative to earlier chemical etching-based experiments by \citet{Blanford_1974}.
The earlier track accumulation rate from \citet{Blanford_1974} is closer to the ACE-derived value (only a factor of $\sim$4 lower); however, as discussed in \citet{Keller_2021}, the TEM measurements are believed to be a more accurate measurement of the track density.
In particular, the TEM measurements are made with relatively thin ($\sim$100$-$150 nm thick) slices of the Apollo lunar sample thereby ensuring a `local' measurement as a function of depth and with respect to the typical track length ($\sim$5$-$15 $\mu$m), while the chemical etching approach used in \citet{Blanford_1974} requires an effective integration over depths of 10$-$15 $\mu$m.
Thus, chemical etching samples a much larger volume which in turn yields an SEP track density that is likely biased to large values relative to the TEM measurements.
Furthermore, while the \citet{Blanford_1974} results are commonly accepted values for chemically etching-based track densities, it was noted even early on that order-of-magnitude discrepancies existed in track density calculations from different groups \citep[e.g., see discussion in][]{Langevin_1977}.

We also note that transmission electron microscope (TEM) measurements can induce fading of SEP-induced damage tracks in minerals \citep[e.g.,][]{Fraundorf_1980, Bradley_1984}. 
Such track fading was particularly noted at electron energies of 100 keV with less pronounced fading at higher energies of 200 keV, where interaction cross sections are typically lower. 
The TEM measurements by \citet{Keller_2021} were conducted at 200 keV electron energies where such fading is not expected to be significant; however, a quantitative analysis of the degree of track fading at 200 keV irradiation has not been fully undertaken.
Nevertheless, we would not expect track fading from 200 keV TEM irradiation to cause the erasure of $\sim$95\% of SEP damage tracks, which is what would be required to explain the difference between the \citet{Keller_2021} results and the ACE in-situ measurements.

% ---------------------

\section{Conclusion}

We have presented a calculation of track-inducing Fe-group SEPs measured at 1 au by the ACE/ULEIS instrument, deriving a flux of $6\times10^5$ cm$^{-2}$ s$^{-1}$ str$^{-1}$.
In comparison, the track accumulation rate determined by laboratory analysis of lunar sample 64455, which was exposed to SEP fluxes on the lunar surface for $\sim$2 Myr, is approximately 25 times lower at $8\times10^3$ cm$^{-2}$ s$^{-1}$ str$^{-1}$.
As discussed above in Section \ref{sec4}, we have considered several possibilities in attempting to explain the difference between the ACE-measured fluxes and those calculated from analysis of lunar sample 64455.
Despite this, no obvious solution for this disagreement is apparent.
While previous work has demonstrated the efficiency of track formation with various minerals at discrete individual energies \citep[e.g.,][]{Price_1973, Szenes_2010}, we would suggest a more thorough investigation.
In particular, an experiment that documented the track registration efficiency across energies spanning the range predicted to induce track formation (i.e., $\sim$0.5-3.0 MeV/nuc) would help to better calibrate the range over which to integrate in-situ measured SEP fluxes.
Such experiments could also examine a variety of mineral phases in order to further constrain any composition-related variations in track registration efficiency.
Additionally, a search for other appropriately suitable lunar samples (whether in the current Apollo collection or to be returned from the upcoming Artemis missions to the Moon) whose SEP-induced track densities over a known lifetime could be compared to those derived from 64455 would provide an additional validation of the results reported in \citet{Keller_2021}.

%%%%%%%%%%%%%%%%%%%%%%%%%%%%%%%%%%%%%%
\newpage

%% ULEIS timeseries

%% ULEIS energy spectrum

%% enstatite Se

%% track density vs depth

\newpage

\begin{acknowledgments}

A. P. gratefully acknowledges support from the NASA New Frontiers Data Analysis Program, grant \#80NSSC18K1557.
A. P. thanks G. Mason for useful discussions on the ACE/ULEIS instrument.

\end{acknowledgments}


\newpage
%\bibliography{/Users/poppe/Projects/Papers/LASP/ref}

%\bibliography{sample63}{}

\begin{thebibliography}{}
\expandafter\ifx\csname natexlab\endcsname\relax\def\natexlab#1{#1}\fi
\providecommand{\url}[1]{\href{#1}{#1}}
\providecommand{\dodoi}[1]{doi:~\href{http://doi.org/#1}{\nolinkurl{#1}}}
\providecommand{\doeprint}[1]{\href{http://ascl.net/#1}{\nolinkurl{http://ascl.net/#1}}}
\providecommand{\doarXiv}[1]{\href{https://arxiv.org/abs/#1}{\nolinkurl{https://arxiv.org/abs/#1}}}

\bibitem[{Afra {et~al.}(2014)Afra, Lang, Bierschenk, Rodriguez, Weber,
  Trautmann, Ewing, Kirby, \& Kluth}]{Afra_2014}
Afra, B., Lang, M., Bierschenk, T., {et~al.} 2014, Nuc. Instr. Meth. Phys. Res.
  B, 326, 126

\bibitem[{Biersack \& Eckstein(1984)}]{Biersack_1984}
Biersack, J.~P., \& Eckstein, W. 1984, Appl. Phys. A, 34, 73

\bibitem[{Blanford {et~al.}(1974)Blanford, Fruland, McKay, \&
  Morrison}]{Blanford_1974}
Blanford, G.~E., Fruland, R.~M., McKay, D.~S., \& Morrison, D.~A. 1974, Proc.
  5th Lunar Conf., 3, 2501

\bibitem[{Bochsler(1987)}]{Bochsler_1987}
Bochsler, P. 1987, Physica Scripta, T18, 55

\bibitem[{Bradley {et~al.}(1984)Bradley, Brownlee, \& Fraundorf}]{Bradley_1984}
Bradley, J.~P., Brownlee, D.~E., \& Fraundorf, P. 1984, Science, 226, 1432,
  \dodoi{10.1126/science.226.4681.1432}

\bibitem[{Bull \& Durrani(1975)}]{Bull_1975}
Bull, R.~K., \& Durrani, S.~A. 1975, Proc. 6th Lunar Sci. Conf., 6, 3619

\bibitem[{Carrasco {et~al.}(2016)Carrasco, Aparicio, Vaquiero, \&
  Gallego}]{Carrasco_2016}
Carrasco, V. M.~S., Aparicio, A. J.~P., Vaquiero, J.~M., \& Gallego, M.~C.
  2016, Solar Phys., 291, 3045, \dodoi{10.1007/s11207-016-0998-7}

\bibitem[{Constantini {et~al.}(1992)Constantini, Brisard, Flament, Meftah,
  Toulemonde, \& Hage-Ali}]{Constantini_1992}
Constantini, J.~M., Brisard, F., Flament, J.~L., {et~al.} 1992, Nuc. Instr.
  Meth. Phys. Res. B, 65, 568

\bibitem[{Crozaz {et~al.}(1972)Crozaz, Drozd, Hohenberg, Hoyt, Ragan, Walker,
  \& Yuhas}]{Crozaz_1972}
Crozaz, G., Drozd, R., Hohenberg, C.~M., {et~al.} 1972, Proc. 3rd Lunar Sci.
  Conf., 3, 2917

\bibitem[{Deca {et~al.}(2015)Deca, Divin, {Lemb\`ege}, Hor{\'a}nyi, Markidis,
  \& Lapenta}]{Deca_2015}
Deca, J., Divin, A., {Lemb\`ege}, B., {et~al.} 2015, J. Geophys. Res.: Space
  Physics, 120, 6443

\bibitem[{Fatemi {et~al.}(2012)Fatemi, Holmstr{\"o}m, \& Futaana}]{Fatemi_2012}
Fatemi, S., Holmstr{\"o}m, M., \& Futaana, Y. 2012, J. Geophys. Res., 117, 1

\bibitem[{Fatemi {et~al.}(2015)Fatemi, Lue, Holmstr{\"o}m, Poppe, Wieser,
  Barabash, \& Delory}]{Fatemi_2015}
Fatemi, S., Lue, C., Holmstr{\"o}m, M., {et~al.} 2015, J. Geophys. Res.: Space
  Physics, 120

\bibitem[{Fleischer {et~al.}(1965)Fleischer, Price, \& Walker}]{Fleischer_1965}
Fleischer, R.~L., Price, P.~B., \& Walker, R.~M. 1965, Annu. Rev. Nucl. Sci.,
  15, 1

\bibitem[{Fleischer {et~al.}(1975)Fleischer, Price, \& Walker}]{Fleischer_1975}
---. 1975, {Nuclear Tracks in Solids: Principles and Applications} ({University
  of California Press})

\bibitem[{Fraundorf {et~al.}(1980)Fraundorf, Flynn, Shirck, \&
  Walker}]{Fraundorf_1980}
Fraundorf, P., Flynn, G.~J., Shirck, J., \& Walker, R.~M. 1980, Proc. Lunar
  Planet. Sci. Conf. 11th, 11, 1235

\bibitem[{Goswami(1981)}]{Goswami_1981}
Goswami, J.~N. 1981, Nature, 293, 124, \dodoi{10.1038/293124a0}

\bibitem[{Keller \& Berger(2014)}]{Keller_2014}
Keller, L.~P., \& Berger, E.~L. 2014, Earth Planets Space, 66,
  \dodoi{10.1186/1880-5981-66-71}

\bibitem[{Keller {et~al.}(2021)Keller, Berger, Zhang, \&
  Christoffersen}]{Keller_2021}
Keller, L.~P., Berger, E.~L., Zhang, S., \& Christoffersen, R. 2021, Meteorit.
  Planet. Sci., 56, 1685, \dodoi{10.1111/maps.13732}

\bibitem[{Keller \& Flynn(2022)}]{Keller_2022}
Keller, L.~P., \& Flynn, G.~J. 2022, Nature Astron., 6, 731,
  \dodoi{10.1038/s41550-022-01647-6}

\bibitem[{Kuchner \& Stark(2010)}]{Kuchner_2010}
Kuchner, M.~J., \& Stark, C.~C. 2010, Astron. J., 140, 1007

\bibitem[{Langevin \& Arnold(1977)}]{Langevin_1977}
Langevin, Y., \& Arnold, J.~R. 1977, Annu. Rev. Earth Planet. Sci., 5, 449

\bibitem[{Liou \& Zook(1999)}]{Liou_1999}
Liou, J.-C., \& Zook, H.~A. 1999, Astron. J., 118, 580

\bibitem[{Liuzzo {et~al.}(2023)Liuzzo, Poppe, Lee, Xu, \&
  Angelopoulos}]{Liuzzo_2023}
Liuzzo, L., Poppe, A.~R., Lee, C.~O., Xu, S., \& Angelopoulos, V. 2023,
  Geophys. Res. Lett., 50, \dodoi{10.1029/2023GL103990}

\bibitem[{Lue {et~al.}(2011)Lue, Futaana, Barabash, Wieser, Holmstr{\"o}m,
  Bhardwaj, Dhanya, \& Wurz}]{Lue_2011}
Lue, C., Futaana, Y., Barabash, S., {et~al.} 2011, Geophys. Res. Lett., 38

\bibitem[{Mason {et~al.}(1998)Mason, Gold, Krimigis, Mazur, Andrews, Daley,
  Dwyer, Heuerman, James, Kennedy, Lefevere, Malcolm, Tossman, \&
  Walpole}]{Mason_1998}
Mason, G.~M., Gold, R.~E., Krimigis, S.~M., {et~al.} 1998, Space Sci. Rev., 86,
  409

\bibitem[{McCracken {et~al.}(2013)McCracken, Beer, Steinhilber, \&
  Abreu}]{McCracken_2013}
McCracken, K.~G., Beer, J., Steinhilber, F., \& Abreu, J. 2013, Solar Phys.,
  286, 609, \dodoi{10.1007/s11207-013-0265-0}

\bibitem[{Meyer(1985)}]{Meyer_1985}
Meyer, J. 1985, Astrophys. J. Supp. Ser., 57, 151

\bibitem[{Mitchell {et~al.}(2008)Mitchell, Halekas, Lin, Frey, Hood, Acu{\~n}a,
  \& Binder}]{Mitchell_2008}
Mitchell, D.~L., Halekas, J.~S., Lin, R.~P., {et~al.} 2008, Icarus, 194, 401

\bibitem[{Muscheler {et~al.}(2016)Muscheler, Adolphi, Herbst, \&
  Nilsson}]{Muscheler_2016}
Muscheler, R., Adolphi, F., Herbst, K., \& Nilsson, A. 2016, Solar Phys., 291,
  3025, \dodoi{10.1007/s11207-016-0969-z}

\bibitem[{Paul \& Fitzgerald(1992)}]{Paul_1992}
Paul, T.~A., \& Fitzgerald, P.~G. 1992, Am. Mineral., 77, 336

\bibitem[{Poppe {et~al.}(2017)Poppe, Halekas, Lue, \& Fatemi}]{Poppe_2017}
Poppe, A.~R., Halekas, J.~S., Lue, C., \& Fatemi, S. 2017, J. Geophys. Res.:
  Planets, 122

\bibitem[{Poppe {et~al.}(2019)Poppe, Lisse, Piquette, Zemcov, Hor{\'a}nyi,
  James, Szalay, Bernardoni, \& Stern}]{Poppe_2019b}
Poppe, A.~R., Lisse, C.~M., Piquette, M., {et~al.} 2019, Astrophys. J. Lett.,
  881, \dodoi{https://doi.org/10.3847/2041-8213/ab322a}

\bibitem[{Price {et~al.}(1973)Price, Lal, Tamhane, \& Perelygin}]{Price_1973}
Price, P.~B., Lal, D., Tamhane, A.~S., \& Perelygin, V.~P. 1973, Earth Plan.
  Sci. Lett., 19, 377

\bibitem[{Price \& {O'Sullivan}(1970)}]{Price_1970}
Price, P.~B., \& {O'Sullivan}, D. 1970, Proc. Apollo 11 Lunar Sci. Conf., 3,
  2351

\bibitem[{Rymzhanov {et~al.}(2019)Rymzhanov, Gorbunov, Medvedev, \&
  Volkov}]{Rymzhanov_2019}
Rymzhanov, R.~A., Gorbunov, S.~A., Medvedev, N., \& Volkov, A.~E. 2019, Nuc.
  Instr. Meth. Phys. Res. B, 440, 25, \dodoi{10.1016/j.nimb.2018.11.034}

\bibitem[{Saito {et~al.}(2012)Saito, Nishino, Fujimoto, Yamamoto, Yokota,
  Tsunakawa, Shibuya, Matsushima, Shimizu, \& Takahashi}]{Saito_2012}
Saito, Y., Nishino, M.~N., Fujimoto, M., {et~al.} 2012, Earth Planets Space,
  64, 83

\bibitem[{Sandford(1986)}]{Sandford_1986}
Sandford, S.~A. 1986, Icarus, 68, 377, \dodoi{10.1016/0019-1035(86)90045-X}

\bibitem[{Seitz {et~al.}(1970)Seitz, Wittels, Maurette, Walker, \&
  Heckman}]{Seitz_1970}
Seitz, M., Wittels, M.~C., Maurette, M., Walker, R.~M., \& Heckman, H. 1970,
  Rad. Effects, 5, 143

\bibitem[{Stone {et~al.}(1998)Stone, Frandsen, Mewaldt, Christian, Margolies,
  Ormes, \& Snow}]{Stone_1998}
Stone, E.~C., Frandsen, A.~M., Mewaldt, R.~A., {et~al.} 1998, Space Sci. Rev.,
  86, \dodoi{10.1007/978-94-011-4762-0_1}

\bibitem[{Szabo {et~al.}(2018)Szabo, Chiba, Biber, Stadlmeyer, Berger, Mayer,
  Mutzke, Doppler, Sauer, Appenroth, Fleig, Foelske-Schmitz, Hutter, Mezger,
  Lammer, Galli, Wurz, \& Aumayr}]{Szabo_2018}
Szabo, P.~S., Chiba, R., Biber, H., {et~al.} 2018, Icarus, 314, 98,
  \dodoi{10.1016/j.icarus.2018.05.028}

\bibitem[{Szenes {et~al.}(2010)Szenes, Kov\'acs, P\'ecz, \&
  Skuratov}]{Szenes_2010}
Szenes, G., Kov\'acs, V.~K., P\'ecz, B., \& Skuratov, V. 2010, Astrophys. J.,
  708, 288, \dodoi{10.1088/0004-637X/708/1/288}

\bibitem[{Thiel {et~al.}(1991)Thiel, Bradley, \& Spohr}]{Thiel_1991}
Thiel, K., Bradley, J.~P., \& Spohr, R. 1991, Nucl. Tracks Radiat. Meas., 19,
  709, \dodoi{10.1016/1359-0189(91)90298-V}

\bibitem[{Usoskin(2023)}]{Usoskin_2023}
Usoskin, I.~G. 2023, Living Rev. Sol. Phys., 20,
  \dodoi{https://doi.org/10.1007/s41116-023-00036-z}

\bibitem[{Usoskin {et~al.}(2016{\natexlab{a}})Usoskin, Gallet, Lopes,
  Kovaltsov, \& Hulot}]{Usoskin_2016}
Usoskin, I.~G., Gallet, Y., Lopes, F., Kovaltsov, G.~A., \& Hulot, G.
  2016{\natexlab{a}}, Astron. Astrophys., 587,
  \dodoi{http://dx.doi.org/10.1051/0004-6361/201527295}

\bibitem[{Usoskin {et~al.}(2016{\natexlab{b}})Usoskin, Kovaltsov, Lockwood,
  Mursula, Owens, \& Solanki}]{Usoskin_2016b}
Usoskin, I.~G., Kovaltsov, G.~A., Lockwood, M., {et~al.} 2016{\natexlab{b}},
  Solar Phys., 291, 2685, \dodoi{10.1007/s11207-015-0838-1}

\bibitem[{{von Rosenvinge} {et~al.}(1995){von Rosenvinge}, Barbier, Karsch,
  Liberman, Madden, Nolan, Reames, Ryan, Singh, Trexel, Winkert, Mason,
  Hamilton, \& Walpole}]{vonRosenvinge_1995}
{von Rosenvinge}, T.~T., Barbier, L.~M., Karsch, J., {et~al.} 1995, Space Sci.
  Rev., 71, 155

\bibitem[{Xu {et~al.}(2017)Xu, Angelopoulos, Wang, Zuo, Wong, \&
  Cui}]{XuX_2017}
Xu, X., Angelopoulos, V., Wang, Y., {et~al.} 2017, Astrophys. J., 849,
  \dodoi{10.3847/1538-4357/aa9186}

\bibitem[{Ziegler {et~al.}(2010)Ziegler, Ziegler, \& Biersack}]{Ziegler_2010}
Ziegler, J.~F., Ziegler, M.~D., \& Biersack, J.~P. 2010, Nuc. Instr. Meth.
  Phys. Res. B, 268, 1818

\bibitem[{Zinner(1980)}]{Zinner_1980}
Zinner, R. 1980, in {The ancient Sun: Fossil record in the Earth, Moon, and
  meteorites}, ed. R.~O. Pepin, J.~A. Eddy, \& R.~B. Merrill ({Pergamon
  Press}), 201--226

\end{thebibliography}
\bibliographystyle{aasjournal}

\end{document}